\documentclass[sigconf,nonacm]{acmart}
\usepackage{listings}
\usepackage[section]{placeins}
\usepackage{array}
\usepackage{vcell}

\usepackage{threeparttable}

\usepackage{float}
\AtBeginDocument{%
  }

\setcopyright{cc}
\setcctype[4.0]{by-nc-nd}

\begin{document}

\title[A Knowledge-Component-Based Methodology for Evaluating AI Assistants]{A Knowledge-Component-Based Methodology\\ for Evaluating AI Assistants}

% SIGCSE Paper 2 title: \title{AI-Enhanced Learning Assessment: Evaluating LLM-Powered Homework Assistance in Introductory Computer Science}


% \author{Laryn Qi}
% \email{larynqi@berkeley.edu}
% \affiliation{%
%   \institution{UC Berkeley EECS}
%   \city{Berkeley}
%   \state{CA}

% \author{J.D. Zamfirescu-Pereira}
% \email{zamfi@berkeley.edu}
% \affiliation{%
%   \institution{UC Berkeley EECS}
%   \city{Berkeley}
%   \state{CA}
% \author{Taehan Kim}
% \email{terry.kim@berkeley.edu}
% \affiliation{%
%   \institution{UC Berkeley EECS}
%   \city{Berkeley}
%   \state{CA}
% \author{Bj\"orn Hartmann}
% \email{bjoern@berkeley.edu}
% \affiliation{%
%   \institution{UC Berkeley EECS}
%   \city{Berkeley}
%   \state{CA}
% \author{John DeNero}
% \email{denero@berkeley.edu}
% \affiliation{%
%   \institution{UC Berkeley EECS}
%   \city{Berkeley}
%   \state{CA}
% \author{Narges Norouzi}
% \email{norouzi@berkeley.edu}
% \affiliation{%
%   \institution{UC Berkeley EECS}
%   \city{Berkeley}
%   \state{CA}

\author{
Laryn Qi \quad J.D. Zamfirescu-Pereira \quad Taehan Kim \\ Bj\"orn Hartmann \quad John DeNero \quad Narges Norouzi \\
\vspace{0.5em}
}
\email{{larynqi, zamfi, terry.kim,bjoern,denero,norouzi}@berkeley.edu}
\affiliation{%
  \institution{\vspace{0.5em}UC Berkeley EECS}
  \city{Berkeley}
  \state{CA}
  \country{USA}
}

\renewcommand{\shortauthors}{Qi, Zamfirescu-Pereira, Kim, Hartmann, DeNero, Norouzi}

\begin{abstract}
%%% JD ABSTRACT %%%
% AI tools are rapidly emerging as a critical component in early CS education---a space now also rich with LLM-based direct student support. Students seem to enjoy using these tools, but there is relatively little formal analysis of these systems' pedagogical effectiveness. In this paper, we propose a method for evaluating a "homework helper" bot for a CS1 course that provides students hints to fix their incomplete code on request. We address three research questions to measure the effectiveness of the assistant's programming hints: \\
% \textbf{RQ1}: \textit{Do the hints help students make progress?} \\
% \textbf{RQ2}: \textit{How effectively do the hints capture problems in student code? }\\
% \textbf{RQ3}: \textit{Are the KCs that students resolve the same as the KCs addressed in the hints?}

% Our method is primarily grounded in identifying knowledge components in homework problems, student code, and helper bot hints. We worked with a team developing one of these helper bots to apply our method to a deployment for 2000+ students across two terms, we show that students are able to resolve problems with their code and approach a working solution more quickly with access to hints, that hints are able to consistently capture the most pressing errors in students' code, and that hints that address a few issues at once rather than a single bug are more likely to lead to direct student progress.

%%% JOHN ABSTRACT %%%
We evaluate an automatic hint generator for CS1 programming assignments powered by GPT-4, a large language model. This system provides natural language guidance about how students can improve their incorrect solutions to short programming exercises. A hint can be requested each time a student fails a test case. Our evaluation addresses three Research Questions: \\
\textbf{RQ1}: \textit{Do the hints help students improve their code?} \\
\textbf{RQ2}: \textit{How effectively do the hints capture problems in student code? }\\
\textbf{RQ3}: \textit{Are the issues that students resolve the same as the issues addressed in the hints?}

To address these research questions quantitatively, we identified a set of fine-grained knowledge components and determined which ones apply to each exercise, incorrect solution, and generated hint. Comparing data from two large CS1 offerings, we found that access to the hints helps students to address problems with their code more quickly, that hints are able to consistently capture the most pressing errors in students' code, and that hints that address a few issues at once rather than a single bug are more likely to lead to direct student progress.
\end{abstract}




\maketitle

\section{Introduction} \label{sec:intro}
Some students struggle to complete programming exercises in CS1 courses even when provided with test cases and test running software. Tutors can potentially help those students make progress by pointing out the specific parts of their incorrect solutions that need revision, as well as the course concepts and techniques relevant to revising them. Large language models (LLMs) have powered a variety of recent programming assistance systems in introductory CS courses~\cite{codeaid,liffiton2023codehelp,stanfordRCT,harvardcs50,gaied}. We use data from the deployment of one of these LLM-based tools that generates hints for CS1 programming exercises that are specific to a student's incorrect solution and failing test cases. % The system we developed, powered by GPT-4, generates hints for CS1 programming exercises that are specific to a student's incorrect solution and failing test cases.

% This paper first gives a brief overview of our tool's deployment and the student workflow.
This paper first gives a brief overview of the dataset and methodology behind our analysis. Then, we dive deeper into students' progression through assignments and quantitatively evaluate the pedagogical effects of LLM-generated hints by answering the following three Research Questions (RQs):

\textbf{RQ1}: \textit{Do the hints help students improve their code?}

\textbf{RQ2}: \textit{How effectively do the hints capture issues in the code?}

% \textbf{RQ3}: \textit{Are students able to apply the hints to make progress?}
\textbf{RQ3}: \textit{Are the issues that students resolve the same as the issues addressed in the hints?}

We evaluate all three of these research questions by identifying and tracking the use of Knowledge Components (KCs), first introduced by Koedinger et al. ~\cite{Koedinger2010TheK}. KCs are fine-grained course concepts and techniques, such as formatting a call expression, using a particular built-in function, updating a variable within a while loop, checking a base case in a recursive function, or accumulating a result and returning it. KCs are a useful tool for evaluating hints because correcting an incorrect solution to a programming exercise often involves correctly applying one or more KCs that are missing from the incorrect solution. Students make progress when the number of missing KCs is reduced. Hints are accurate when they correctly address the missing KCs. When students apply missing KCs referenced in a hint, that's evidence that they are applying the hint to make progress.

We evaluated the RQs quantitatively by analyzing students' responses to programming exercises and the system's hints for two homework assignments that were similar in two large offerings of a CS1 course. %We found that, with assistant access, a greater proportion of code revisions reduced missing KCs (RQ1), indicating that students more often made progress using the hints.
We found that, with assistant access, students made fewer reverse-progress/no-progress revisions where the number of missing KCs in their code either went up or did not change, respectively---upwards of a 50\% drop in these unproductive submissions for the majority of problems. We also found high overlap between the KCs missing from incorrect solutions and the KCs referenced in generated hints, indicating that hints were often accurate (RQ2)---over 85\% of \textit{all} hints addressed a high-priority missing KC. Finally, we found evidence that the KCs students resolve were often the ones addressed in the hints, indicating that some students were able to understand and act on the hints (RQ3). Furthermore, hints that addressed \textit{multiple} missing KCs led to direct student progress more often---more than half the time for hints that address multiple KCs.

\section{Background \& Related Work}
\subsection{LLM-Based Tools in CS Education}
In large introductory Computer Science (CS) courses, students who are new to CS get stuck often on programming assignments ~\cite{cs1-repeaters, cs1-cant-program} and there tends to be more demand for staff help than available staff office hours ~\cite{inp-vs-rem-OH}---%two problems that are 
a problem that is alleviated with Artificial Intelligence (AI) assistants. The rise of Large Language Models (LLMs) has opened the door for the straightforward development of AI educational tools that provide on-demand feedback with high availability. As a result, development of AI tools for education has skyrocketed in the past year. Researchers in academia and industry are experimenting with incorporating State-Of-The-Art (SOTA) LLMs into their classrooms and products.

% In just the past year, many LLM-based coding assistants have been developed and deployed in introductory CS courses ~\cite{gaied, harvardAI, stanfordRCT, codeaid, liffiton2023codehelp}.% In our case, we have developed a homework assistant powered by GPT-4 ~\cite{gpt4} that looks at students' incomplete code, the output of student code on a set of test cases, along with the past few failed student attempts and responds with feedback and hints for improvement.

In academia, many large introductory CS courses are using LLMs to provide support to students at scale in the form of debugging help on programming assignments ~\cite{gaied, harvardAI, stanfordRCT, codeaid, liffiton2023codehelp, program-repair}, answering online forum questions ~\cite{edLLM, harvardAI}, and generating course materials ~\cite{sarsa2022automatic}. Generally, these tools seem well-received by students and save time for both students and teaching staff. While preliminary usage statistics and surveys suggest that students enjoy using these tools, there is far less analysis on how these tools are affecting students' learning.

In the private sector, the programming capability of LLMs is being leveraged across many industries. In research, some companies are exploring tuning foundation models to specialize in the task of code generation, such as Codex ~\cite{chen2021evaluating}, while others are building end-to-end agents on top of LLMs that can act as autonomous software engineers like Devin~\cite{devin}%\footnote{Devin, \url{https://www.cognition.ai/blog/introducing-devin}}
. Developer-facing companies are releasing coding companions such as Copilot~\cite{copilot} that can help programmers by generating template code, investigating bugs, or writing documentation and test cases. Educational organizations are developing AI tutors and Teaching Assistants (TAs) like Khanmigo~\cite{khanmigo} which are capable of guiding students to answers and taking care of teachers' busy work. As with similar projects in academia, these industry products have seen wide adoption and generally positive or optimistic reception.

\subsection{Knowledge Component Framework}
Traditionally, mastery of KCs is measured via a statistical procedure (Bayesian Knowledge Tracing) ~\cite{bkt, cmu-unproductive-struggle}, but more recent work has shown the capacity of deep learning approaches toward both generation of KCs and evaluation of KC mastery. In 2023, a multilayer perceptron was shown to perform well on automatic KC generation and KC mastery evaluation tasks ~\cite{kc-finder}. Most recently, SOTA LLMs are able to effectively generate a set of KCs and extract KCs from student code in CS1 ~\cite{niousha2024kcs}. We make use of this capability of LLMs and extend it to also extract KCs from the hints generated by the assistant.% As other related work in detecting unproductive struggle does ~\cite{cmu-unproductive-struggle}, we also utilize KCs as a proxy for measuring student progress.

While recent related work ~\cite{wang2024bridging} has found success in using topic-agnostic classifications to categorize student errors, we decided to use (topic-specific) KCs to align the concepts missing in student code one-to-one with the concepts addressed by hints.

More details about our KC framework and extraction can be found in the following section \S\ref{sec:dataset}.

% \section{Method: Notes on Design \& Deployment}
\section{Method: Dataset \& Knowledge Component Extraction} \label{sec:dataset}

We apply our method to the data of \textbf{61A-Bot} ~\cite{gaied}, one these LLM-based helper tools. Starting in Spring 2024, the assistant was fully deployed to 872 enrolled students in CS 61A, the introductory CS course at UC Berkeley.

\subsection{Extracting Knowledge Components Missing in Code}
We extract KCs from student code across two semesters: Spring 2023 and Spring 2024. We treat Spring 2023 (\textit{before} the assistant's release) data as a baseline for comparison with Spring 2024 (\textit{after} the assistant's release). These code samples are intermediate, incomplete student code checkpoints that are logged on a server after each autograder run.

We can find which KCs each student is missing as they progress through a homework problem by looking at their code checkpoints. We pass the student's code, a list of human-generated KCs% (see Appendix ~\ref{app:kc-list})
, and some metadata % (see Appendix ~\ref{app:rq1-prompt})
to GPT-4 and ask it to determine which KCs the student is missing from the provided list and output them in order of most important to least important missing KC (additionally, we instruct the LLM to output its reasoning for its chosen KCs in a Chain-of-Thought ~\cite{wei2023chainofthought}-inspired manner). This approach is a variation on previous LLM-based KC extraction work, which would ask the LLM to \textit{generate} missing KCs from scratch rather than \textit{classify} missing KCs from a provided list. We found that GPT-4 performed better if given the list of possible KCs compared to expert labels. Our human-generated KCs range from relatively high-level "mini-concepts" (e.g. "Function Objects versus Function Calls", "Return versus Print") to common errors students make on these problems (e.g. "Return Early in Loop", "Forget While Loop Update"). This list of KCs, which includes short descriptions/definitions for each KC, effectively serves as a database for the LLM to select from, reducing the possibility for hallucination. For a single homework, we manually defined anywhere from 30-40 KCs across 7-9 categories (e.g. "Control", "Functions", "Recursion"), depending on the specific homework topics.

Additionally, we considered pre-assigning the list of possible KCs at the problem-level instead of at the assignment-level (each assignment consists of multiple problems). While this approach would limit the cases where the LLM hallucinates about KCs that are not relevant to the current problem, it would also limit the LLM's capacity to measure the progression of creative alternate solutions that students come up with.

% \laryn{(1a) need accuracy comparison with human labels (full list)?}

To measure the accuracy of GPT-4 at this task of extracting missing KCs from code (with a reference KC list provided), we sampled 63 pairs of student code checkpoints and corresponding LLM outputs. Then, two experts (co-authors who are also TAs for the course) made a binary classification of whether they agreed with the LLM-generated missing KC list. The two experts had an inter-rater agreement of 93\% with a Cohen's $k$ of $0.759$ % \jd{we should use one of the measures of IRR, not just list percentage agreement, to account for baseline agreement given ground truth distribution (which may be more than 50\%).}
on a small, jointly labeled sub-sample of 14 queries (agreed on 13/14). %\jd{also, how did we get 93\% agreement on 10 labels?}.
Overall, the experts agreed with \textbf{85\%} of the LLM-generated missing KC list.% In a sample of 63 such queries to GPT-4, 85\% of the LLM-generated missing KC lists were considered reasonable by experts (co-authors on this paper and also TAs for the course). The two experts agreed on 93\% of a small, mutually labeled dataset. 

%Following a similar process as in \S~\ref{sec:rq1}, 

\subsection{Extracting Knowledge Components Addressed in Hints}
For extracting KCs addressed by assistant hints, we only look at the Spring 2024, the term where the tool was fully deployed. We follow a similar process as we did with student code but now for the assistant's hints: we pass the hint, a list of human-generated KCs, and some metadata %(see Appendix ~\ref{app:rq2-prompt})
to GPT-4 and ask it to determine which KCs the hint addresses from the provided list and, this time, output them in any order (as well as its reasoning).

To measure the accuracy of GPT-4 at this task of identifying the KCs addressed by a hint (with a reference KC list provided), we sampled 63 pairs of hints and corresponding LLM outputs---for the purpose of expert lableling, we had the LLM output only one KC addressed by the hint at a time. Then, two experts (co-authors who are also TAs for the course) made a binary classification of whether they agreed that the LLM-generated KCs also existed in the hint. The two experts had an inter-rater agreement of 92\% with a Cohen's $k$ of $0.629$ on a small, jointly labeled sub-sample of 13 queries (agreed on 12/13). Overall, the experts found that \textbf{87\%} of the time, GPT-4 correctly classified whether a particular KC existed in the hint.

% In a sample of 62 similar queries, 87\% of the time, the LLM correctly classified whether a KC existed in the hint compared to expert labels (labeled by co-authors on this paper who are also TAs for the course). The two experts agreed on 92\% of a small, mutually labeled dataset. \laryn{(1b) need accuracy comparison with human labels (full list)?}

\section{Results} \label{sec:results}
% In an effort to quantitatively evaluate the pedagogical effects of the hints, we pose 3 research questions (RQs):

% 1. Do the hints help students improve their code?

% 2. How effectively do the hints capture issues in the code?

% 3. Were students able to use the hints to improve their code?

% As a framework for evaluating these 3 RQs, we ground our analysis in \textbf{Knowledge Components} (KCs), first introduced by Koedinger et al. ~\cite{Koedinger2010TheK}. KCs represent course topics that students gain mastery of by completing assignments. They provide a finer-grained metric of student progression throughout an assignment which we use to measure students' abilities to grok and apply the assistant hints as well as the assistant's ability to address relevant issues in students' understanding. Previous work by Niousha et al. ~\cite{niousha2024kcs} has shown the capacity of state of the art (SOTA) LLMs to effectively extract KCs from code in CS1. We make use of this capability and extend it to also extract KCs from the hints generated by the assistant.

\subsection{RQ1: Do the hints help students improve their code?} \label{sec:rq1}
% \begin{figure}[H]
\begin{figure}[!htb]
    \centering
    \includegraphics[width=0.85\linewidth]{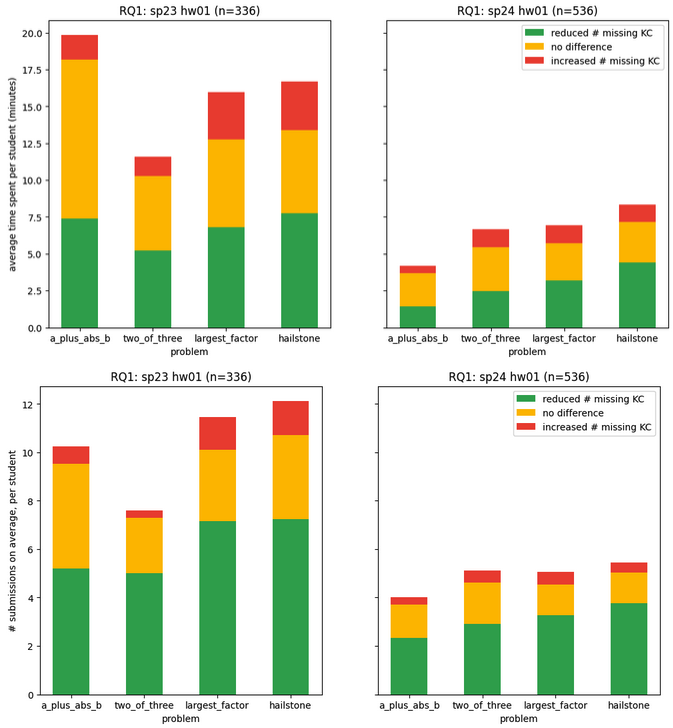}
    \caption{Sp23 (Left) vs. Sp24 (Right) HW1 Time Spent \& Number of Submissions Colored by Student Progression. Upper plots show average time spent per student. Lower plots show average number of submissions per student.}
    \label{fig:rq1-hw1}
\end{figure}
\begin{figure}[!htb]
% \begin{figure}[H]
    \centering
    \includegraphics[width=0.85\linewidth]{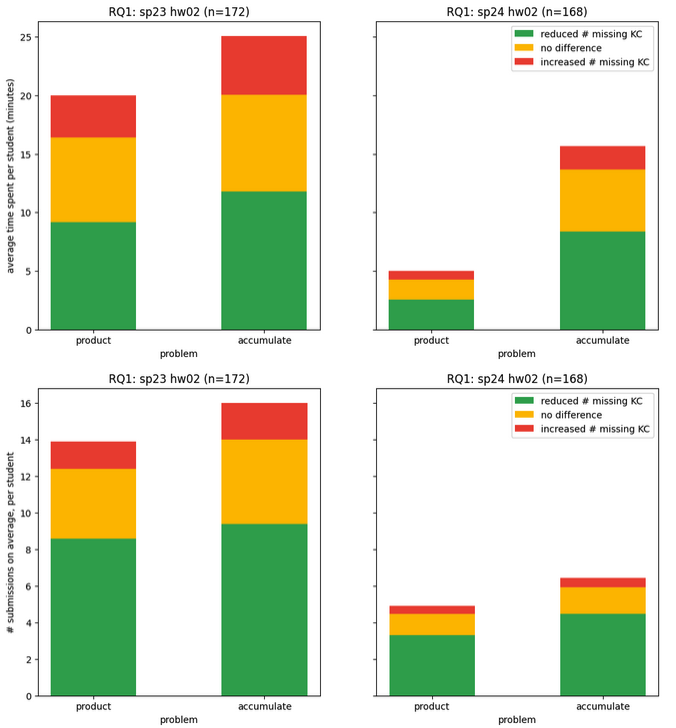}
    \caption{Sp23 (Left) vs. Sp24 (Right) HW2 Time Spent \& Number of Submissions Colored by Student Progression. Upper plots show average time spent per student. Lower plots show average number of submissions per student.}
    \label{fig:rq1-hw2}
\end{figure}
\begin{figure}[!htb]
% \begin{figure}[H]
    \centering
    \includegraphics[width=0.85\linewidth]{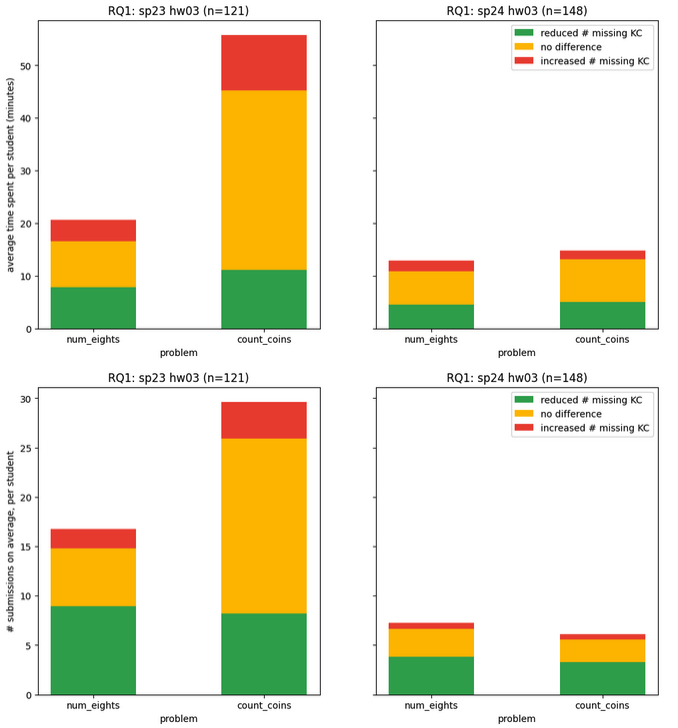}
    \caption{Sp23 (Left) vs. Sp24 (Right) HW3 Time Spent \& Number of Submissions Colored by Student Progression. Upper plots show average time spent per student. Lower plots show average number of submissions per student.}
    \label{fig:rq1-hw3}
\end{figure}

With the dataset of missing KCs lists associated with each code checkpoint, we look at time-adjacent pairs of checkpoints and compare whether the number of missing KCs decreases, increases, or stays the same. With this categorization, we can visualize how students' progressions through assignments change with the introduction of the assistant, as shown in Figures ~\ref{fig:rq1-hw1}, ~\ref{fig:rq1-hw2}, and ~\ref{fig:rq1-hw3}. Specifically, we calculate the average \textit{time spent} per student and average \textit{number of submissions} made per student in each of these three categories. In our plots, green regions correspond to checkpoint pairs where the number of missing KCs decreases; red regions correspond to pairs where the number of missing KCs increases, and yellow regions correspond to pairs where the number of missing KCs stays the same.

From Spring 2023 (no-assistant) to Spring 2024 (with-assistant), we observe drops across the board both in time spent and the number of submissions associated with each color region%, which aligns with our completion time findings in \S~\ref{sec:completion-time}
. This drop is upwards of 50\% for the majority of problems including \texttt{a\_plus\_abs\_b}, \texttt{largest\_factor}, \texttt{hailstone}, \texttt{product}, and \texttt{coun\\t\_coins}. Spending less time in yellow and red regions may be attributable to the assistant's hints pointing out issues in their code more quickly than the students themselves can. The reduction in the green region is less intuitive but may be explained by Spring 2024 students making \textit{more progress \textbf{per green submission}} compared to Spring 2023 students. In Spring 2024, each green submission reduces the number of missing KCs by $1.42$ ($SD=0.71$) on average, while in Spring 2023, each green submission reduces the number of missing KCs by $1.28$ ($SD=0.56$) on average ($p < 0.000001$, $r = 0.107$, $t$-statistic$ = 4.93$, $dof = 2116$).% \jd{add r, t-statistic, dof here -- assume this is an unpaired t-test -- see https://yatani.jp/teaching/doku.php?id=hcistats:ttest for more info} %A $t$-test tells us that this difference in sample means is statistically significant. Specifically, assuming equal population standard deviations in Spring 2023 and Spring 2024, there is less than a $0.00001\%$ chance of this occurring if the two population means were the same. \jd{just say $p < 0.00001$.}

%\jd{is this difference statistically significant? what are the SDs or variances on these means?}

\begin{figure}[!htb]
% \begin{figure}[H]
    \centering
    \includegraphics[width=0.85\linewidth]{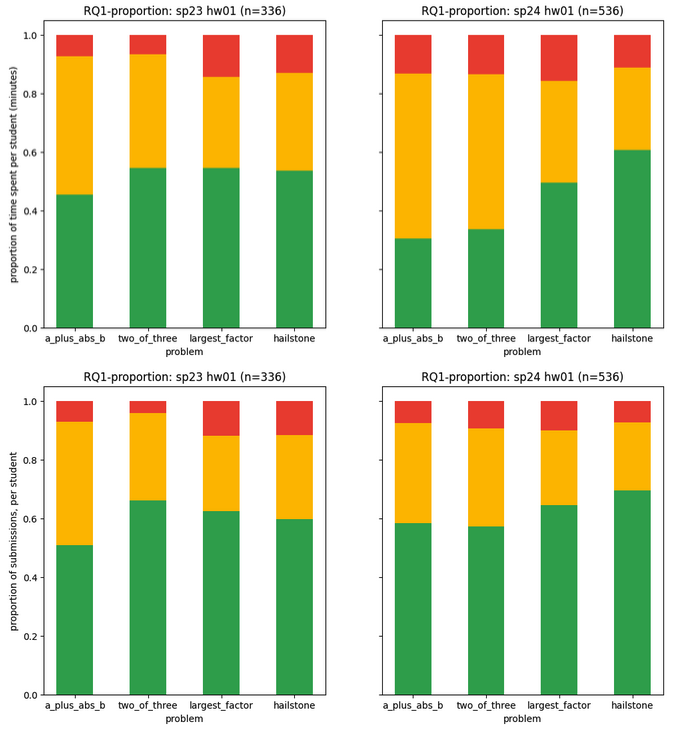}
    \caption{Sp23 (Left) vs. Sp24 (Right) HW1 \textit{Proportion} of Time Spent \& Number of Submissions Colored by Student Progression. Upper plots show average time spent per student. Lower plots show average number of submissions per student.}
    \label{fig:rq1-hw1-prop}
\end{figure}

\begin{figure}[!htb]
% \begin{figure}[H]
    \centering
    \includegraphics[width=0.85\linewidth]{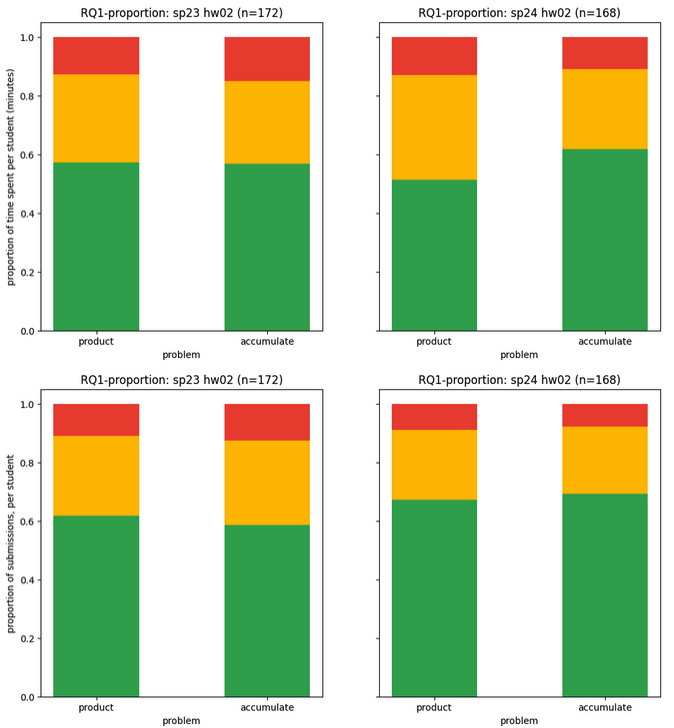}
    \caption{Sp23 (Left) vs. Sp24 (Right) HW2 \textit{Proportion} of Time Spent \& Number of Submissions Colored by Student Progression. Upper plots show average time spent per student. Lower plots show average number of submissions per student.}
    \label{fig:rq1-hw2-prop}
\end{figure}

\begin{figure}[!htb]
% \begin{figure}[H]
    \centering
    \includegraphics[width=0.85\linewidth]{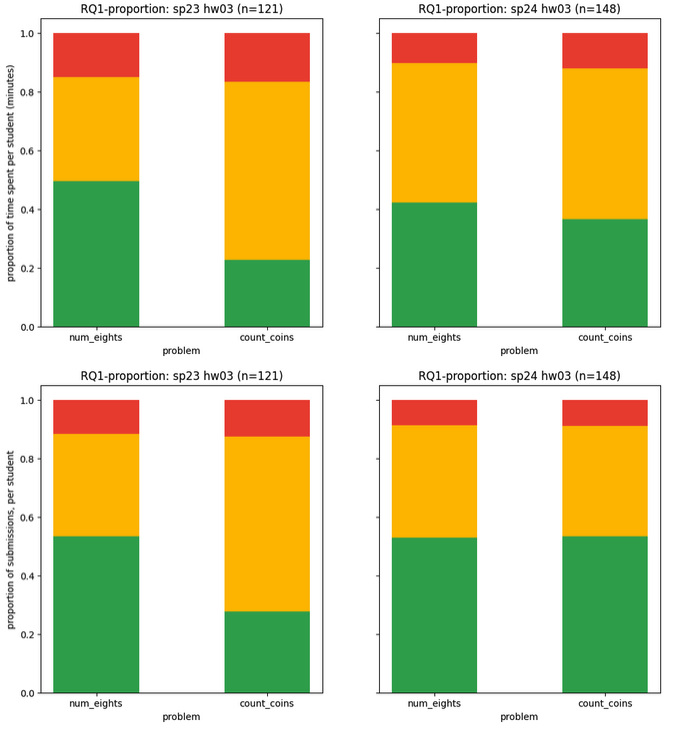}
    \caption{Sp23 (Left) vs. Sp24 (Right) HW3 \textit{Proportion} of Time Spent \& Number of Submissions Colored by Student Progression. Upper plots show average time spent per student. Lower plots show average number of submissions per student.}
    \label{fig:rq1-hw3-prop}
\end{figure}

If we look at the time spent in each color category \textit{proportionally} (see Figures ~\ref{fig:rq1-hw1-prop}, ~\ref{fig:rq1-hw2-prop}, and ~\ref{fig:rq1-hw3-prop}), specifically in terms of the number of submissions (rather than time), we can see that the distribution of categories is mostly consistent from Spring 2023 to Spring 2024 (with some cases---e.g. HW3---showing significant increases in green regions). This suggests that students' general progression through problems remains similar and is merely expedited by the assistant in Spring 2024, i.e., students inevitably encounter and still struggle to resolve the same bugs and issues regardless of hints but overall move at a faster pace when the assistant is available.% See Appendix ~\ref{app:rq1-plots} for plots of additional homeworks.
% \jd{i really like this interpretation. is there any other data we can draw on to corroborate this finding, though? 
% (if not yet, let's brainstorm for SIGCSE)}

One natural question that arises with this color categorization is, "Are there any patterns in these hints that make them good (green) or bad (yellow/red)?" Qualitatively, we've noticed that in some red hint cases (e.g., \texttt{Your two\_of\_three function is still not correct. The problem is that you're not always getting the two smallest numbers. For example, if i is 1, j is 2, and k is 3, your current function would return 1*1 + 2*2 = 5, which is correct. But if i is 3, j is 1, and k is 2, your function would return 1*1 + 1*1 = 2, which is incorrect. You need to find a way to exclude the largest number from the calculation.}), a clear fault in the hint is that it simply points out a mistake (and provides a failure case), but does not suggest a solution or fix. However, it does not seem to be a consistent property of red (nor yellow) hints. Similarly, for green hints, the successful application of the hint seems to be heavily dependent on the student's:

(a) Understanding of course topics and the CS1 terminology in order to digest what the hint is telling them to do

and

(b) Understanding of their own code, which is often incomplete yet they choose to run the autograder anyway just to see what happens.

Overall, it seems that we need more context in order to differentiate between green, yellow, and red hints in most cases.

\subsection{RQ2: How effectively do the hints capture issues in the code?} \label{sec:rq2}
% \laryn{(3) add intuition for why splitting by kc list length}
Next, we explore which KCs the \textit{hints} address and look at the overlap between the KCs the hint discusses and the missing KCs from code as a measure of how effectively the assistant hints are able to capture issues in student code. With the list of KCs addressed by each hint along with the list of missing KCs for each student code checkpoint from \S~\ref{sec:rq1}, we can measure the overlap. Specifically, we calculate the proportion of hints that address \textit{at least 1 out of the top-3} missing KCs in the student's code. We use this metric of "overlap with top-3 missing KCs" because our goal with the assistant is not to have its hints point out \textit{every} single bug and misunderstanding the student has---instead, we are interested in whether the assistant's hints are addressing the \textit{most pressing} missing KCs in the student code, similar to how a human tutor would walk through a problem with a student in office hours. \textbf{Over 85\% of all hints} given for homeworks 1, 2, and 3 address a top-3 critical missing KC.

Furthermore, we choose to group hints by \textit{the number of KCs addressed by the hint} (i.e., KC hint list length, hint KC-richness) as we're interested in exploring whether leveraging the capability of the assistant to enumerate multiple issues in the student's code in a single hint, allowing the student to come back re-read as they make edits to their code (a unique benefit of written versus spoken tutoring ~\cite{gaied}), leads to better student progress.

%As seen in Figures ~\ref{fig:rq2-hw1} and ~\ref{fig:rq2-hw2}, 

\begin{figure}[!htb]
% \begin{figure}[H]
    \centering
    \includegraphics[width=0.75\linewidth]{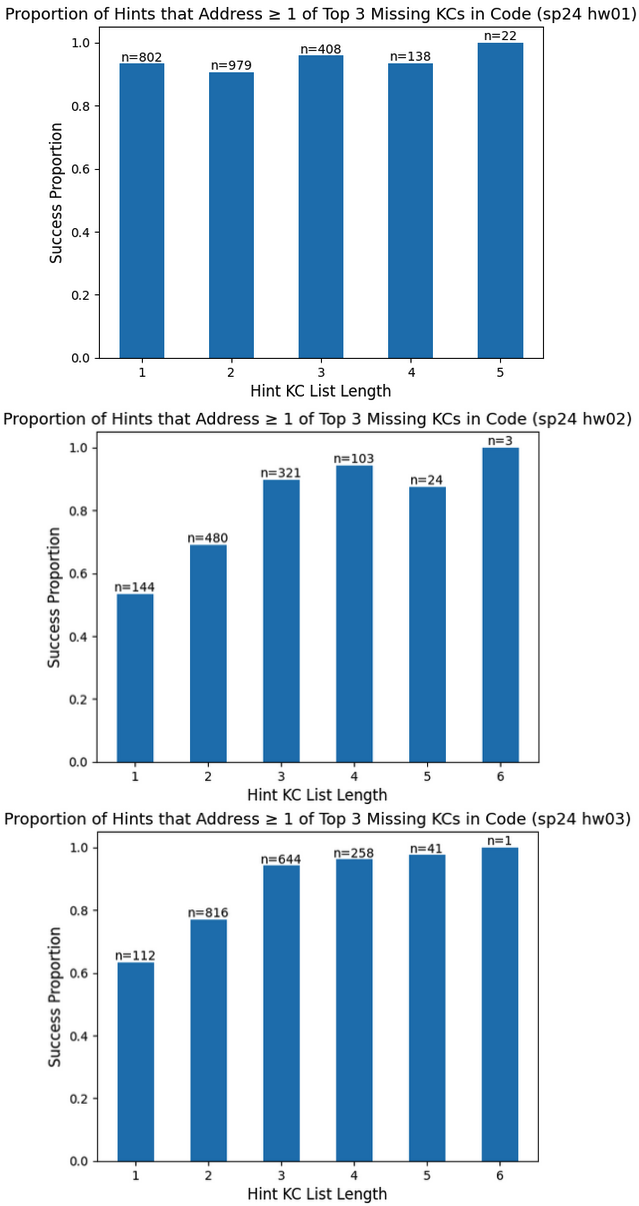}
    \caption{Sp24 HWs 1,2,3 Proportion of Hints that Address $\geq$ 1 of Top 3 Missing KCs in Code Grouped By Hint KC List Length.}
    \label{fig:rq2}
\end{figure}

In homework 1 (see Figure ~\ref{fig:rq2}) we notice that over 85\% of hints fulfill this overlap criterion for \textit{all} hint KC list lengths. This is surprising as we would expect hints that cover a larger number of KCs to have a higher likelihood of also covering one of the top-3 missing KCs in the student's code, but the success proportion appears nearly uniform over all hint KC list lengths. Another implication of this hint of KC-richness independence is that the LLM is quite good at recognizing and weighing the importance of various issues in student code and ensuring that the \textit{key} issues are included in its response---possibly an unintended but welcome side-effect of including a \hspace{0.05cm}\texttt{Limit your response to a sentence or two at most} instruction in the prompt to the assistant.

% \laryn{add here}

% \jd{unclear why this is broken down by KC list length?}

For later homeworks (i.e., 2 \& 3), we do not observe as strong an overlap between KCs addressed by hint and missing KCs in the code, specifically with KC-sparse hints (i.e., shorter hint KC list length). We speculate this is because our few-shot expert examples were limited to samples from homework 1, which leads to less accurate KC classifications in the hints and the code checkpoints. The effect of this lower accuracy is exacerbated in the low-data settings of hints that only address one or two KCs. Naturally, there is an upward slope that tapers off past a hint KC list length of $3$ which is most likely due to the fact that covering more KCs leads to higher probability of addressing a KC that is relevant and missing in the student's code just by chance.  % See Appendix ~\ref{app:rq2-plots} for plots of additional homeworks.

% \begin{figure}[!htb]
% % \begin{figure}[H]
%     \centering
%     \includegraphics[width=0.8\linewidth]{figs/rq2/rq2-hw1.png}
%     \caption{Sp24 HW1 Proportion of Hints that Address $\geq$ 1 of Top 3 Missing KCs in Code Grouped By Hint KC List Length.}
%     \label{fig:rq2-hw1}
% \end{figure}

% \begin{figure}[!htb]
% % \begin{figure}[H]
%     \centering
%     \includegraphics[width=0.8\linewidth]{figs/rq2/rq2-hw2.png}
%     \caption{Sp24 HW2 Proportion of Hints that Address $\geq$ 1 of Top 3 Missing KCs in Code Grouped By Hint KC List Length.}
%     \label{fig:rq2-hw2}
% \end{figure}

% \begin{figure}[!htb]
% % \begin{figure}[H]
%     \centering
%     \includegraphics[width=0.8\linewidth]{figs/rq2/rq2-hw3.png}
%     \caption{Sp24 HW3 Proportion of Hints that Address $\geq$ 1 of Top 3 Missing KCs in Code Grouped By Hint KC List Length}
%     \label{fig:rq2-hw3}
% \end{figure}

% \laryn{(2) add paragraph for qualitative analysis of good vs bad hints?}

\subsection{RQ3: Are the issues that students resolve the same as the issues addressed in the hints?} \label{sec:rq3}
\begin{figure}[!htb]
% \begin{figure}[H]
    \centering
    \includegraphics[width=0.75\linewidth]{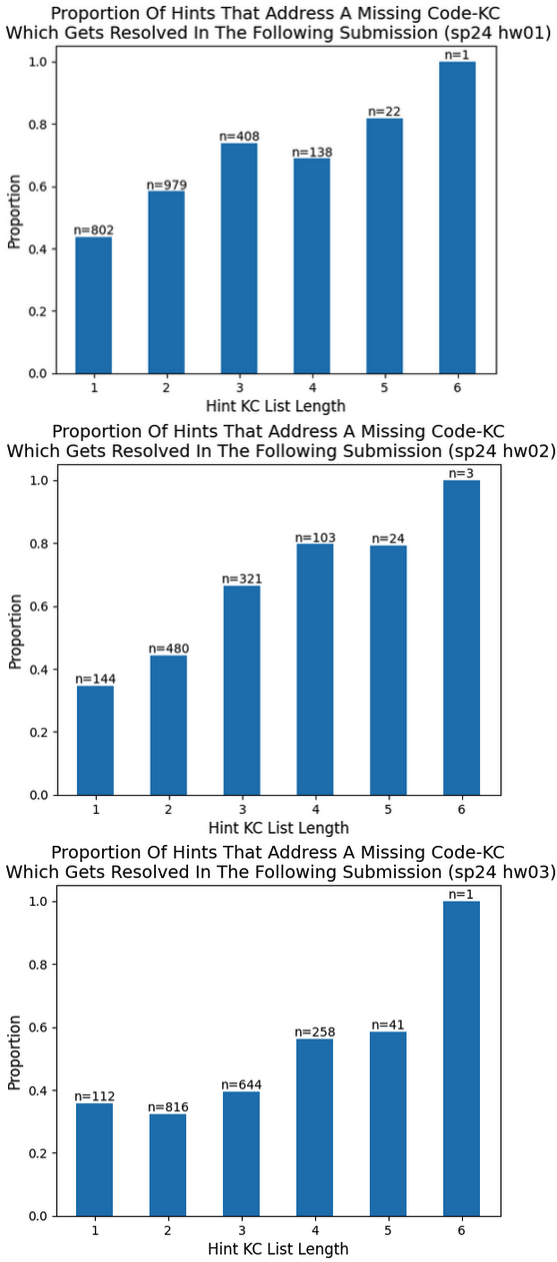}
    \caption{Sp24 HWs 1,2,3 Proportion of Hints that Address a Subsequently Resolved KC Grouped By Hint KC List Length.}
    \label{fig:rq3}
\end{figure}

% \subsection{RQ3: Are students able to apply the hints to make progress?} \label{sec:rq3}
Finally, to measure how well students are able to apply the hints to improve their code, we can check whether the KCs that are missing in the student's code at checkpoint $i$ are no longer missing at checkpoint $i + 1$ (i.e., the resolved KCs) are the same as the KCs addressed in the hint given in response to the the checkpoint $i$ code. % We start by using the same calculation as done before in \S~\ref{sec:rq2}. Then, we introduce the same coloring scheme as in \S~\ref{sec:rq1} to indicate what proportion of hints that address a top-3 KC \textit{are also} successfully applied by students to make direct progress in their following code submission.
Specifically, for each hint KC list length, we count the proportion of hints that address at least one missing KC that is resolved in the following submission. As we see in Figure ~\ref{fig:rq3}%Figures ~\ref{fig:rq3-hw1}, ~\ref{fig:rq3-hw2} and ~\ref{fig:rq3-hw3}% (for homeworks 1 and 2)% (and Appendix ~\ref{app:rq3-plots})
, hints that cover a greater number of KCs is more likely to lead to direct student progress. This suggests that steering the LLM to cover 3 or 4 issues in a single hint could lead to student progress more often than honing in on one bug---a bug, which, in reality, could be a minor issue when compared to the student's general approach to the problem, for example. In contrast to the previous result in \S~\ref{sec:rq2}, we see the number of KCs addressed by the hint playing a key role in students' ability to apply at least \textit{one} of the suggestions in the hint. While KC-sparse hints may be able to address core issues in student code consistently, there's a better chance of the student successfully applying \textit{some} part of the hint to make progress when the hint is KC-rich. This effect is less apparent in homework 3 (Figure ~\ref{fig:rq3-hw3}) which we suspect may be due to the higher conceptual difficulty of later homeworks, in which case the hint's ability to explain a concept well likely has a higher correlation with immediate student progress than KC-richness.% See Appendix ~\ref{app:green} \& ~\ref{app:yellow} for samples of green \& yellow submission pairs respectively.
\section{Next Steps \& Conclusion} \label{sec:next-steps}
To extend the work we did in \S~\ref{sec:results}, we plan to perform additional \textit{time-series} analysis of student submissions to gain an extra dimension of insight into how individual students are progressing through assignments differently with and without access to the assistant. Moreover, one clear drawback of the KC-based progression categorization we've developed is that the \textit{number} of missing KCs does not always accurately measure how much progress a student is making. For example, a code submission with a slew of "Off-by-One" and "Forget While Loop Update" errors could realistically be much closer to a correct solution than a recursive function that has a single "Incorrect Recursive Recombination Step" missing KC associated with it. So, collecting more data on this approach's validity (or shortcomings) will prove valuable.

% \begin{figure}[H]
% \begin{figure}[!htb]
%     \centering
%     \includegraphics[width=0.75\linewidth]{figs/rq3/rq3-hw1.png}
%     \caption{Sp24 HW1 Proportion of Hints that Address a Subsequently Resolved KC (Grouped By Hint KC List Length).}
%     \label{fig:rq3-hw1}
% \end{figure}

% \begin{figure}[!htb]
% % \begin{figure}[H]
%     \centering
%     \includegraphics[width=0.75\linewidth]{figs/rq3/rq3-hw2.png}
%     \caption{Sp24 HW2 Proportion of Hints that Address a Subsequently Resolved KC (Grouped By Hint KC List Length).}
%     \label{fig:rq3-hw2}
% \end{figure}

% \begin{figure}[!htb]
% % \begin{figure}[H]
%     \centering
%     \includegraphics[width=0.75\linewidth]{figs/rq3/rq3-hw3.png}
%     \caption{Sp24 HW3 Proportion of Hints that Address a Subsequently Resolved KC (Grouped By Hint KC List Length).}
%     \label{fig:rq3-hw3}
% \end{figure}

Inspired by recent work in model distillation ~\cite{chen2024personalised}, we experimented with inserting a "judge" LLM in the middle of our pipeline that would take in an LLM-generated list of KCs as input and either validate it or make changes to it. While we found the "judge" outputs to be, on average, as good as the initial LLM outputs (when evaluated by co-authors), we are hopeful that future iterations on the "judge" prompt that include few-shot expert examples or using an ensemble of diverse LLMs for evaluation (i.e., a "jury") ~\cite{verga2024replacing} will result in significant improvements in LLM-generated KC accuracy.

% \section{Conclusion} \label{sec:conclusion}

% Next steps / challanges to improve:

% -- dialog, or at least buttons like "say more" or "i dont get it" etc.

% -- Worked examples

% -- ...?

% Short-term plan: address worked examples, include prior context for specific question in every new query, offer more button options than just "ok"?

% Longer-term plan: assess performance of students who rely heavily on tool, compare students with tool access to those without , etc.

% We first reported here on preliminary findings from an early deployment of a GPT-4-based interactive programming support tool for introductory CS courses. We found a number of successes, identified a few challenges and potential pitfalls, and reflected on solutions and paths toward more complete automated support for introductory CS students.

%We first reported here on the deployment of an LLM-based interactive programming support tool for a large introductory CS course. Additionally, we explored 
\begin{minipage}{\linewidth}
We reported here on using \textbf{knowledge components} as a proxy for student progress and found that students follow the same general progression trajectories through problems---only faster with the assistant. Analyzing the hints themselves revealed that the LLM \textit{is} able to decipher important issues in student code at a high rate and that hints that cover a number of issues rather than a single bug are more likely to lead to direct student progress.

% We discovered a number of successes, identified a few challenges and pitfalls, and reflected on solutions and paths toward a more pedagogically robust LLM assistant for introductory CS students.

\begin{acks}
% test
This work was made possible by a few generous sources of support: an Inclusion Research Award from Google, and support for the assistant's use of Azure's OpenAI API by Microsoft.

% This work and deployment were supported by a generous sponsorship of GPT-4 use by Microsoft.

\end{acks}
\end{minipage}


\bibliographystyle{ACM-Reference-Format}
\bibliography{ref,csbot}


\begin{thebibliography}{25}

%%% the Web address.

\ifx \showCODEN    \undefined \def \showCODEN     #1{\unskip}     \fi
\ifx \showDOI      \undefined \def \showDOI       #1{#1}\fi
\ifx \showISBNx    \undefined \def \showISBNx     #1{\unskip}     \fi
\ifx \showISBNxiii \undefined \def \showISBNxiii  #1{\unskip}     \fi
\ifx \showISSN     \undefined \def \showISSN      #1{\unskip}     \fi
\ifx \showLCCN     \undefined \def \showLCCN      #1{\unskip}     \fi
\ifx \shownote     \undefined \def \shownote      #1{#1}          \fi
\ifx \showarticletitle \undefined \def \showarticletitle #1{#1}   \fi
\ifx \showURL      \undefined \def \showURL       {\relax}        \fi
% invisible to TeX
\providecommand\bibfield[2]{#2}
\providecommand\bibinfo[2]{#2}
\providecommand\natexlab[1]{#1}
\providecommand\showeprint[2][]{arXiv:#2}

\bibitem[Academy(2023)]%
        {khanmigo}
\bibfield{author}{\bibinfo{person}{Khan Academy}.} \bibinfo{year}{2023}\natexlab{}.
\newblock \bibinfo{title}{Khanmigo}.
\newblock
\newblock
\urldef\tempurl%
\url{https://www.khanmigo.ai}
\showURL{%
\tempurl}


\bibitem[Chen et~al\mbox{.}(2024)]%
        {chen2024personalised}
\bibfield{author}{\bibinfo{person}{Hailin Chen}, \bibinfo{person}{Amrita Saha}, \bibinfo{person}{Steven Hoi}, {and} \bibinfo{person}{Shafiq Joty}.} \bibinfo{year}{2024}\natexlab{}.
\newblock \bibinfo{title}{Personalised Distillation: Empowering Open-Sourced LLMs with Adaptive Learning for Code Generation}.
\newblock
\newblock
\showeprint[arxiv]{2310.18628}~[cs.CL]


\bibitem[Chen et~al\mbox{.}(2021)]%
        {chen2021evaluating}
\bibfield{author}{\bibinfo{person}{Mark Chen}, \bibinfo{person}{Jerry Tworek}, \bibinfo{person}{Heewoo Jun}, \bibinfo{person}{Qiming Yuan}, \bibinfo{person}{Henrique Ponde de~Oliveira Pinto}, \bibinfo{person}{Jared Kaplan}, \bibinfo{person}{Harri Edwards}, \bibinfo{person}{Yuri Burda}, \bibinfo{person}{Nicholas Joseph}, \bibinfo{person}{Greg Brockman}, {et~al\mbox{.}}} \bibinfo{year}{2021}\natexlab{}.
\newblock \showarticletitle{Evaluating large language models trained on code}.
\newblock \bibinfo{journal}{\emph{arXiv preprint arXiv:2107.03374}} (\bibinfo{year}{2021}).
\newblock


\bibitem[Cognition(2024)]%
        {devin}
\bibfield{author}{\bibinfo{person}{Cognition}.} \bibinfo{year}{2024}\natexlab{}.
\newblock \bibinfo{title}{Devin}.
\newblock
\newblock
\urldef\tempurl%
\url{https://www.cognition.ai/blog/introducing-devin}
\showURL{%
\tempurl}


\bibitem[Corbett and Anderson(1994)]%
        {bkt}
\bibfield{author}{\bibinfo{person}{Albert Corbett, A.T.} {and} \bibinfo{person}{John Anderson, R.}} \bibinfo{year}{1994}\natexlab{}.
\newblock \showarticletitle{Knowledge tracing: Modeling the acquisition of procedural knowledge}.
\newblock \bibinfo{journal}{\emph{User Modeling and User-Adapted Interaction}} (\bibinfo{year}{1994}).
\newblock


\bibitem[Donlevy(2023)]%
        {harvardcs50}
\bibfield{author}{\bibinfo{person}{Katherine Donlevy}.} \bibinfo{year}{2023}\natexlab{}.
\newblock \bibinfo{title}{Harvard to roll out AI professors in flagship coding class for fall semester.}
\newblock
\newblock
\urldef\tempurl%
\url{https://nypost.com/2023/06/30/harvard-to-roll-out-ai-professors-in-flagship-coding-class-for-fall-semester/}
\showURL{%
\tempurl}


\bibitem[Fancsali et~al\mbox{.}(2020)]%
        {cmu-unproductive-struggle}
\bibfield{author}{\bibinfo{person}{Stephen~E. Fancsali}, \bibinfo{person}{Kenneth Holstein}, \bibinfo{person}{Michael Sandbothe}, \bibinfo{person}{Steven Ritter}, \bibinfo{person}{Bruce~M. McLaren}, {and} \bibinfo{person}{Vincent Aleven}.} \bibinfo{year}{2020}\natexlab{}.
\newblock \showarticletitle{Towards Practical Detection of Unproductive Struggle}. In \bibinfo{booktitle}{\emph{Artificial Intelligence in Education}}, \bibfield{editor}{\bibinfo{person}{Ig~Ibert Bittencourt}, \bibinfo{person}{Mutlu Cukurova}, \bibinfo{person}{Kasia Muldner}, \bibinfo{person}{Rose Luckin}, {and} \bibinfo{person}{Eva Mill{\'a}n}} (Eds.). \bibinfo{publisher}{Springer International Publishing}, \bibinfo{address}{Cham}, \bibinfo{pages}{92--97}.
\newblock
\showISBNx{978-3-030-52240-7}


\bibitem[Gao et~al\mbox{.}(2022)]%
        {inp-vs-rem-OH}
\bibfield{author}{\bibinfo{person}{Zhikai Gao}, \bibinfo{person}{Sarah Heckman}, {and} \bibinfo{person}{Collin Lynch}.} \bibinfo{year}{2022}\natexlab{}.
\newblock \showarticletitle{Who Uses Office Hours? A Comparison of In-Person and Virtual Office Hours Utilization}. In \bibinfo{booktitle}{\emph{Proceedings of the 53rd ACM Technical Symposium on Computer Science Education - Volume 1}} (Providence, RI, USA) \emph{(\bibinfo{series}{SIGCSE 2022})}. \bibinfo{publisher}{Association for Computing Machinery}, \bibinfo{address}{New York, NY, USA}, \bibinfo{pages}{300–306}.
\newblock
\showISBNx{9781450390705}
\urldef\tempurl%
\url{https://doi.org/10.1145/3478431.3499334}
\showDOI{\tempurl}


\bibitem[GitHub(2024)]%
        {copilot}
\bibfield{author}{\bibinfo{person}{GitHub}.} \bibinfo{year}{2024}\natexlab{}.
\newblock \bibinfo{title}{Copilot}.
\newblock
\newblock
\urldef\tempurl%
\url{https://github.com/features/copilot}
\showURL{%
\tempurl}


\bibitem[Kazemitabaar et~al\mbox{.}(2024)]%
        {codeaid}
\bibfield{author}{\bibinfo{person}{Majeed Kazemitabaar}, \bibinfo{person}{Runlong Ye}, \bibinfo{person}{Xiaoning Wang}, \bibinfo{person}{Austin~Zachary Henley}, \bibinfo{person}{Paul Denny}, \bibinfo{person}{Michelle Craig}, {and} \bibinfo{person}{Tovi Grossman}.} \bibinfo{year}{2024}\natexlab{}.
\newblock \showarticletitle{CodeAid: Evaluating a Classroom Deployment of an LLM-based Programming Assistant that Balances Student and Educator Needs}. In \bibinfo{booktitle}{\emph{Proceedings of the CHI Conference on Human Factors in Computing Systems}} (Honolulu, HI, USA) \emph{(\bibinfo{series}{CHI '24})}. \bibinfo{publisher}{Association for Computing Machinery}, \bibinfo{address}{New York, NY, USA}, Article \bibinfo{articleno}{650}, \bibinfo{numpages}{20}~pages.
\newblock
\showISBNx{9798400703300}
\urldef\tempurl%
\url{https://doi.org/10.1145/3613904.3642773}
\showDOI{\tempurl}


\bibitem[Koedinger et~al\mbox{.}(2010)]%
        {Koedinger2010TheK}
\bibfield{author}{\bibinfo{person}{K. Koedinger}, \bibinfo{person}{Albert~T. Corbett}, {and} \bibinfo{person}{Charles~A. Perfetti}.} \bibinfo{year}{2010}\natexlab{}.
\newblock \showarticletitle{The Knowledge-Learning-Instruction (KLI) Framework: Toward Bridging the Science-Practice Chasm to Enhance Robust Student Learning}.
\newblock
\urldef\tempurl%
\url{https://api.semanticscholar.org/CorpusID:15732814}
\showURL{%
\tempurl}


\bibitem[Koutcheme et~al\mbox{.}(2023)]%
        {program-repair}
\bibfield{author}{\bibinfo{person}{Charles Koutcheme}, \bibinfo{person}{Sami Sarsa}, \bibinfo{person}{Juho Leinonen}, \bibinfo{person}{Arto Hellas}, {and} \bibinfo{person}{Paul Denny}.} \bibinfo{year}{2023}\natexlab{}.
\newblock \showarticletitle{Automated Program Repair Using Generative Models for Code Infilling}. In \bibinfo{booktitle}{\emph{Artificial Intelligence in Education}}, \bibfield{editor}{\bibinfo{person}{Ning Wang}, \bibinfo{person}{Genaro Rebolledo-Mendez}, \bibinfo{person}{Noboru Matsuda}, \bibinfo{person}{Olga~C. Santos}, {and} \bibinfo{person}{Vania Dimitrova}} (Eds.). \bibinfo{publisher}{Springer Nature Switzerland}, \bibinfo{address}{Cham}, \bibinfo{pages}{798--803}.
\newblock
\showISBNx{978-3-031-36272-9}


\bibitem[Liffiton et~al\mbox{.}(2023)]%
        {liffiton2023codehelp}
\bibfield{author}{\bibinfo{person}{Mark Liffiton}, \bibinfo{person}{Brad Sheese}, \bibinfo{person}{Jaromir Savelka}, {and} \bibinfo{person}{Paul Denny}.} \bibinfo{year}{2023}\natexlab{}.
\newblock \bibinfo{title}{CodeHelp: Using Large Language Models with Guardrails for Scalable Support in Programming Classes}.
\newblock
\newblock
\showeprint[arxiv]{2308.06921}~[cs.CY]


\bibitem[Lister et~al\mbox{.}(2004)]%
        {cs1-cant-program}
\bibfield{author}{\bibinfo{person}{Raymond Lister}, \bibinfo{person}{Elizabeth~S. Adams}, \bibinfo{person}{Sue Fitzgerald}, \bibinfo{person}{William Fone}, \bibinfo{person}{John Hamer}, \bibinfo{person}{Morten Lindholm}, \bibinfo{person}{Robert McCartney}, \bibinfo{person}{Jan~Erik Mostr\"{o}m}, \bibinfo{person}{Kate Sanders}, \bibinfo{person}{Otto Sepp\"{a}l\"{a}}, \bibinfo{person}{Beth Simon}, {and} \bibinfo{person}{Lynda Thomas}.} \bibinfo{year}{2004}\natexlab{}.
\newblock \showarticletitle{A multi-national study of reading and tracing skills in novice programmers}. In \bibinfo{booktitle}{\emph{Working Group Reports from ITiCSE on Innovation and Technology in Computer Science Education}} (Leeds, United Kingdom) \emph{(\bibinfo{series}{ITiCSE-WGR '04})}. \bibinfo{publisher}{Association for Computing Machinery}, \bibinfo{address}{New York, NY, USA}, \bibinfo{pages}{119–150}.
\newblock
\showISBNx{9781450377942}
\urldef\tempurl%
\url{https://doi.org/10.1145/1044550.1041673}
\showDOI{\tempurl}


\bibitem[Liu et~al\mbox{.}(2024)]%
        {harvardAI}
\bibfield{author}{\bibinfo{person}{Rongxin Liu}, \bibinfo{person}{Carter Zenke}, \bibinfo{person}{Charlie Liu}, \bibinfo{person}{Andrew Holmes}, \bibinfo{person}{Patrick Thornton}, {and} \bibinfo{person}{David~J. Malan}.} \bibinfo{year}{2024}\natexlab{}.
\newblock \showarticletitle{Teaching CS50 with AI: Leveraging Generative Artificial Intelligence in Computer Science Education}. In \bibinfo{booktitle}{\emph{Proceedings of the 55th ACM Technical Symposium on Computer Science Education V. 2}} (Portland, OR, USA) \emph{(\bibinfo{series}{SIGCSE 2024})}. \bibinfo{publisher}{Association for Computing Machinery}, \bibinfo{address}{New York, NY, USA}, \bibinfo{pages}{1927}.
\newblock
\showISBNx{9798400704246}
\urldef\tempurl%
\url{https://doi.org/10.1145/3626253.3635427}
\showDOI{\tempurl}


\bibitem[Marco-Galindo et~al\mbox{.}(2022)]%
        {cs1-repeaters}
\bibfield{author}{\bibinfo{person}{María-Jesús Marco-Galindo}, \bibinfo{person}{Julià Minguillón}, \bibinfo{person}{David García-Solórzano}, {and} \bibinfo{person}{Teresa Sancho-Vinuesa}.} \bibinfo{year}{2022}\natexlab{}.
\newblock \showarticletitle{Why Do CS1 Students Become Repeaters?}
\newblock \bibinfo{journal}{\emph{IEEE Revista Iberoamericana de Tecnologias del Aprendizaje}} \bibinfo{volume}{17}, \bibinfo{number}{3} (\bibinfo{year}{2022}), \bibinfo{pages}{245--253}.
\newblock
\urldef\tempurl%
\url{https://doi.org/10.1109/RITA.2022.3191288}
\showDOI{\tempurl}


\bibitem[Mitra et~al\mbox{.}(2024)]%
        {edLLM}
\bibfield{author}{\bibinfo{person}{Chancharik Mitra}, \bibinfo{person}{Mihran Miroyan}, \bibinfo{person}{Rishi Jain}, \bibinfo{person}{Vedant Kumud}, \bibinfo{person}{Gireeja Ranade}, {and} \bibinfo{person}{Narges Norouzi}.} \bibinfo{year}{2024}\natexlab{}.
\newblock \showarticletitle{Elevating Learning Experiences: Leveraging Large Language Models as Student-Facing Assistants in Discussion Forums}. In \bibinfo{booktitle}{\emph{Proceedings of the 55th ACM Technical Symposium on Computer Science Education V. 2}} (Portland, OR, USA) \emph{(\bibinfo{series}{SIGCSE 2024})}. \bibinfo{publisher}{Association for Computing Machinery}, \bibinfo{address}{New York, NY, USA}, \bibinfo{pages}{1752–1753}.
\newblock
\showISBNx{9798400704246}
\urldef\tempurl%
\url{https://doi.org/10.1145/3626253.3635609}
\showDOI{\tempurl}


\bibitem[Niousha et~al\mbox{.}(2024)]%
        {niousha2024kcs}
\bibfield{author}{\bibinfo{person}{Rose Niousha}, \bibinfo{person}{Muntasir Hoq}, \bibinfo{person}{Bita Akram}, {and} \bibinfo{person}{Narges Norouzi}.} \bibinfo{year}{2024}\natexlab{}.
\newblock \showarticletitle{Use of Large Language Models for Extracting Knowledge Components in CS1 Programming Exercises}. In \bibinfo{booktitle}{\emph{Proceedings of the 55th ACM Technical Symposium on Computer Science Education V. 2}} (Portland, OR, USA) \emph{(\bibinfo{series}{SIGCSE 2024})}. \bibinfo{publisher}{Association for Computing Machinery}, \bibinfo{address}{New York, NY, USA}, \bibinfo{pages}{1762–1763}.
\newblock
\showISBNx{9798400704246}
\urldef\tempurl%
\url{https://doi.org/10.1145/3626253.3635592}
\showDOI{\tempurl}


\bibitem[Sarsa et~al\mbox{.}(2022)]%
        {sarsa2022automatic}
\bibfield{author}{\bibinfo{person}{Sami Sarsa}, \bibinfo{person}{Paul Denny}, \bibinfo{person}{Arto Hellas}, {and} \bibinfo{person}{Juho Leinonen}.} \bibinfo{year}{2022}\natexlab{}.
\newblock \showarticletitle{Automatic generation of programming exercises and code explanations using large language models}. In \bibinfo{booktitle}{\emph{Proceedings of the 2022 ACM Conference on International Computing Education Research-Volume 1}}. \bibinfo{pages}{27--43}.
\newblock


\bibitem[Shi et~al\mbox{.}(2023)]%
        {kc-finder}
\bibfield{author}{\bibinfo{person}{Yang Shi}, \bibinfo{person}{Robin Schmucker}, \bibinfo{person}{Min Chi}, \bibinfo{person}{Tiffany Barnes}, {and} \bibinfo{person}{Thomas Price}.} \bibinfo{year}{2023}\natexlab{}.
\newblock \showarticletitle{KC-Finder: Automated Knowledge Component Discovery for Programming Problems}. In \bibinfo{booktitle}{\emph{Proceedings of the 16th International Conference on Educational Data Mining}}, \bibfield{editor}{\bibinfo{person}{Mingyu Feng}, \bibinfo{person}{Tanja KÃ¤ser}, {and} \bibinfo{person}{Partha Talukdar}} (Eds.). \bibinfo{publisher}{International Educational Data Mining Society}, \bibinfo{address}{Bengaluru, India}, \bibinfo{pages}{28--39}.
\newblock
\showISBNx{978-1-7336736-4-8}
\urldef\tempurl%
\url{https://doi.org/10.5281/zenodo.8115671}
\showDOI{\tempurl}


\bibitem[Verga et~al\mbox{.}(2024)]%
        {verga2024replacing}
\bibfield{author}{\bibinfo{person}{Pat Verga}, \bibinfo{person}{Sebastian Hofstatter}, \bibinfo{person}{Sophia Althammer}, \bibinfo{person}{Yixuan Su}, \bibinfo{person}{Aleksandra Piktus}, \bibinfo{person}{Arkady Arkhangorodsky}, \bibinfo{person}{Minjie Xu}, \bibinfo{person}{Naomi White}, {and} \bibinfo{person}{Patrick Lewis}.} \bibinfo{year}{2024}\natexlab{}.
\newblock \bibinfo{title}{Replacing Judges with Juries: Evaluating LLM Generations with a Panel of Diverse Models}.
\newblock
\newblock
\showeprint[arxiv]{2404.18796}~[cs.CL]


\bibitem[Wang et~al\mbox{.}(2024b)]%
        {wang2024bridging}
\bibfield{author}{\bibinfo{person}{Rose~E. Wang}, \bibinfo{person}{Qingyang Zhang}, \bibinfo{person}{Carly Robinson}, \bibinfo{person}{Susanna Loeb}, {and} \bibinfo{person}{Dorottya Demszky}.} \bibinfo{year}{2024}\natexlab{b}.
\newblock \bibinfo{title}{Bridging the Novice-Expert Gap via Models of Decision-Making: A Case Study on Remediating Math Mistakes}.
\newblock
\newblock
\showeprint[arxiv]{2310.10648}~[cs.CL]


\bibitem[Wang et~al\mbox{.}(2024a)]%
        {stanfordRCT}
\bibfield{author}{\bibinfo{person}{Sierra Wang}, \bibinfo{person}{John Mitchell}, {and} \bibinfo{person}{Chris Piech}.} \bibinfo{year}{2024}\natexlab{a}.
\newblock \showarticletitle{A Large Scale RCT on Effective Error Messages in CS1}. In \bibinfo{booktitle}{\emph{Proceedings of the 55th ACM Technical Symposium on Computer Science Education V. 1}} (Portland, OR, USA) \emph{(\bibinfo{series}{SIGCSE 2024})}. \bibinfo{publisher}{Association for Computing Machinery}, \bibinfo{address}{New York, NY, USA}, \bibinfo{pages}{1395–1401}.
\newblock
\showISBNx{9798400704239}
\urldef\tempurl%
\url{https://doi.org/10.1145/3626252.3630764}
\showDOI{\tempurl}


\bibitem[Wei et~al\mbox{.}(2023)]%
        {wei2023chainofthought}
\bibfield{author}{\bibinfo{person}{Jason Wei}, \bibinfo{person}{Xuezhi Wang}, \bibinfo{person}{Dale Schuurmans}, \bibinfo{person}{Maarten Bosma}, \bibinfo{person}{Brian Ichter}, \bibinfo{person}{Fei Xia}, \bibinfo{person}{Ed Chi}, \bibinfo{person}{Quoc Le}, {and} \bibinfo{person}{Denny Zhou}.} \bibinfo{year}{2023}\natexlab{}.
\newblock \bibinfo{title}{Chain-of-Thought Prompting Elicits Reasoning in Large Language Models}.
\newblock
\newblock
\showeprint[arxiv]{2201.11903}~[cs.CL]


\bibitem[Zamfirescu-Pereira et~al\mbox{.}(2023)]%
        {gaied}
\bibfield{author}{\bibinfo{person}{J.D. Zamfirescu-Pereira}, \bibinfo{person}{Laryn Qi}, \bibinfo{person}{Bjoern Hartmann}, \bibinfo{person}{John DeNero}, {and} \bibinfo{person}{Narges Norouzi}.} \bibinfo{year}{2023}\natexlab{}.
\newblock \bibinfo{title}{Conversational Programming with LLM-Powered Interactive Support in an Introductory Computer Science Course}.
\newblock
\newblock
\urldef\tempurl%
\url{https://gaied.org/neurips2023/files/32/32_paper.pdf}
\showURL{%
\tempurl}


\end{thebibliography}

\end{document}